\documentclass[sigconf]{acmart}

\usepackage[linesnumbered,vlined,ruled,commentsnumbered,noend]{algorithm2e}
\SetKw{Def}{def}
\SetKw{Continue}{continue}
\SetKw{Break}{break}
\SetKwComment{Comment}{$//$}{}
\SetKwFunction{schedule}{schedule}
\SetKwFunction{commit}{commit}

\AtBeginDocument{%
  \providecommand\BibTeX{{%
    \normalfont B\kern-0.5em{\scshape i\kern-0.25em b}\kern-0.8em\TeX}}}

\copyrightyear{2022}
\acmYear{2022}
\setcopyright{acmcopyright}\acmConference[ICSE '22]{44th International Conference on Software Engineering}{May 21--29, 2022}{Pittsburgh, PA, USA}
\acmBooktitle{44th International Conference on Software Engineering (ICSE '22), May 21--29, 2022, Pittsburgh, PA, USA}
\acmPrice{15.00}
\acmDOI{10.1145/3510003.3510086}
\acmISBN{978-1-4503-9221-1/22/05}

\usepackage{listings, xcolor}

\definecolor{verylightgray}{rgb}{.97,.97,.97}

\lstdefinelanguage{Solidity}{
	keywords=[1]{anonymous, assembly, assert, balance, break, call, callcode, case, catch, class, constant, continue, constructor, contract, debugger, default, delegatecall, delete, do, else, emit, event, experimental, export, external, false, finally, for, function, gas, if, implements, import, in, indexed, instanceof, interface, internal, is, length, library, log0, log1, log2, log3, log4, memory, modifier, new, payable, pragma, private, protected, public, pure, push, require, return, returns, revert, selfdestruct, send, solidity, storage, struct, suicide, super, switch, then, this, throw, transfer, true, try, typeof, using, value, view, while, with, addmod, ecrecover, keccak256, mulmod, ripemd160, sha256, sha3}, %
	keywordstyle=[1]\color{blue}\bfseries,
	keywords=[2]{address, bool, byte, bytes, bytes1, bytes2, bytes3, bytes4, bytes5, bytes6, bytes7, bytes8, bytes9, bytes10, bytes11, bytes12, bytes13, bytes14, bytes15, bytes16, bytes17, bytes18, bytes19, bytes20, bytes21, bytes22, bytes23, bytes24, bytes25, bytes26, bytes27, bytes28, bytes29, bytes30, bytes31, bytes32, enum, int, int8, int16, int24, int32, int40, int48, int56, int64, int72, int80, int88, int96, int104, int112, int120, int128, int136, int144, int152, int160, int168, int176, int184, int192, int200, int208, int216, int224, int232, int240, int248, int256, mapping, string, uint, uint8, uint16, uint24, uint32, uint40, uint48, uint56, uint64, uint72, uint80, uint88, uint96, uint104, uint112, uint120, uint128, uint136, uint144, uint152, uint160, uint168, uint176, uint184, uint192, uint200, uint208, uint216, uint224, uint232, uint240, uint248, uint256, var, void, ether, finney, szabo, wei, days, hours, minutes, seconds, weeks, years},	%
	keywordstyle=[2]\color{teal}\bfseries,
	keywords=[3]{block, blockhash, coinbase, difficulty, gaslimit, number, timestamp, msg, data, gas, sender, sig, value, now, tx, gasprice, origin},	%
	keywordstyle=[3]\color{violet}\bfseries,
	identifierstyle=\color{black},
	sensitive=false,
	comment=[l]{//},
	morecomment=[s]{/*}{*/},
	commentstyle=\color{gray}\ttfamily,
	stringstyle=\color{red}\ttfamily,
	morestring=[b]',
	morestring=[b]"
}

\begin{document}

\title{Utilizing Parallelism in Smart Contracts on Decentralized Blockchains by Taming Application-Inherent Conflicts}

\author{Péter Garamvölgyi}
\email{peter.garamvolgyi@confluxnetwork.org}
\affiliation{%
  \institution{Shanghai Tree-Graph Blockchain Research Institute}
  \city{Shanghai}
  \country{China}
}

\author{Yuxi Liu}
\email{yuxi.liu@duke.edu}
\affiliation{%
  \institution{Duke University}
  \city{Durham}
  \state{North Carolina}
  \country{USA}
}
\authornote{Work done while employed at Shanghai Qi Zhi Institute.}

\author{Dong Zhou}
\email{dongz@mail.tsinghua.edu.cn}
\affiliation{%
  \institution{Tsinghua University}
  \city{Beijing}
  \country{China}
}
\affiliation{%
  \institution{Shanghai Qi Zhi Institute}
  \city{Shanghai}
  \country{China}
}

\author{Fan Long}
\email{fanl@cs.toronto.edu}
\affiliation{%
  \institution{University of Toronto}
  \city{Toronto}
  \country{Canada}
}
\affiliation{%
  \institution{Shanghai Tree-Graph Blockchain Research Institute}
  \city{Shanghai}
  \country{China}
}

\author{Ming Wu}
\email{ming.wu@confluxnetwork.org}
\affiliation{%
  \institution{Shanghai Tree-Graph Blockchain Research Institute}
  \city{Shanghai}
  \country{China}
}

\begin{abstract}

Traditional public blockchain systems typically had very limited transaction throughput because of the bottleneck of the consensus protocol itself.
With recent advances in consensus technology, the performance limit has been greatly lifted, typically to thousands of transactions per second.
With this, transaction execution has become a new performance bottleneck.
Exploiting parallelism in transaction execution is a clear and direct way to address this and to further increase transaction throughput.
Although some recent literature introduced concurrency control mechanisms to execute smart contract transactions in parallel, the reported speedup that they can achieve is far from ideal.
The main reason is that the proposed parallel execution mechanisms cannot effectively deal with the conflicts inherent in many blockchain applications.

In this work, we thoroughly study the historical transaction execution traces in Ethereum.
We observe that application-inherent conflicts are the major factors that limit the exploitable parallelism during execution.
We propose to use partitioned counters and special commutative instructions to break up the application conflict chains in order to maximize the potential speedup.
When we evaluated the maximum parallel speedup achievable, these techniques doubled this limit to an 18x overall speedup compared to serial execution, thus approaching the optimum.
We also propose OCC-DA, an optimistic concurrency control scheduler with deterministic aborts, which makes it possible to use OCC scheduling in public blockchain settings.

\end{abstract}



\maketitle

\section{Introduction} \label{introduction}

The technical challenge of scaling permissionless blockchains has been a hot research topic for the last few years. With various scaling solutions, be it Ethereum 2.0's sharding~\cite{eth2} or Conflux's Tree-Graph ledger structure~\cite{conflux}, the consensus mechanism ceases to be the performance bottleneck. While disk I/O, network bandwidth, and transaction execution are all possible sources of contention, transaction execution is arguably the most challenging one to address.

Distributed ledgers that follow the account model originally introduced by Ethereum are designed to reach consensus on a sequence of transactions, then process them serially. As a result, current protocols and their implementations are unable to make use of multiple threads on multi-core processors during this execution step. Given the dependencies between transactions through their accesses to a shared data structure called the state tree, the first challenge is to understand how much speedup we can potentially achieve by executing them in parallel. Then, the second challenge is to design a parallel scheduler with sufficient determinism so that nodes can reach consensus.

To understand the degree of parallelism that can be utilized in existing transaction workloads, this paper empirically studied a period of historical Ethereum transactions. Taking state access traces (perfect information), transaction gas costs, and the degree of parallelism of computing resources (e.g., 32 threads) as inputs, we constructed an optimal schedule for each block, then compared its execution time to that of serial execution. Our major findings include:

\begin{enumerate}
\item The overall speedup achievable is limited at about 4x compared to serial execution. While there are many blocks whose execution scales with the number of threads, a large portion of blocks performs significantly worse. These results are consistent with previous works~\cite{saraph2019empirical, reijsbergen2020exploiting}.
\item Most blocks are bottlenecked on a single chain of dependent transactions that need to be executed serially and thus dominate the overall execution time. 
\item A manual inspection of the bottleneck transactions shows that most of them conflict on a single counter or array. From the application's perspective, most bottleneck transactions can be classified into one of three categories: token distribution, collectibles, and decentralized finance.
\end{enumerate}

The empirical study results suggest that, instead of optimizing scheduler implementations, our primary focus should be on eliminating these common sources of contention in smart contracts.
In this paper, we present three independent techniques for eliminating the aforementioned bottlenecks.
Orthogonal to these techniques, we also present a novel scheduling framework called \emph{optimistic concurrency control with deterministic aborts (OCC-DA).}
Parallel schedulers that follow this framework can comply with the stringent determinism requirements of distributed consensus.

The first, simplest approach to eliminating bottlenecks is to use multiple sender addresses.
By manually dividing a set of transactions from a single sender to multiple disjoint sets of transactions, many common bottleneck patterns can be eliminated.

The second approach is to use \textit{partitioned counters}, similar to \textit{sloppy counters}, originally introduced by Boyd-Wickizer et al.~\cite{boyd2010analysis} for the Linux kernel. In this approach, we maintain several sub-counters, the sum of which constitutes the value of the original counter. Writes are routed to and operate on different sub-counters based on some attribute, e.g., the sender's address. This way, partitioned counters reduce the probability that any two writing transactions will conflict.

The third approach to addressing bottlenecks is to bypass avoidable conflicts arising from commutative updates on the virtual machine level. Two transactions that both increment a counter but do not use its original value are semantically commutative. However, under the current Ethereum Virtual Machine semantics such increments are translated into a read (\texttt{SLOAD}) and a write (\texttt{SSTORE}) instruction which will lead to read-write conflicts. We propose a new instruction called \texttt{CADD} (commutative add). Two transactions that only have \texttt{CADD} operations but no other reads and writes on a given state entry are not considered conflicting. Increments are applied during transaction commit serially.

Our evaluations suggest that these approaches can raise the amount of speedup achievable to 18x or more, making it approach the optimal case where all transaction dependencies are ignored.

We also note that the non-determinism that is characteristic of parallel execution might prevent blockchain nodes from reaching consensus. A set of incentives for \emph{good} behavior (i.e., following the protocol) and dis-incentives for \emph{bad} behavior (i.e., attacking or misusing the protocol) is an essential part of permissionless blockchains. Ethereum and similar systems offer no incentive to write smart contracts or pack blocks in a way that improves transaction parallelizability. The number of conflicts and/or transaction aborts is a metric of parallelizability that the incentive layer could use to assign financial rewards and penalties. However, under traditional approaches like optimistic concurrency control (OCC)~\cite{kung1981optimistic}, even if we enforce a deterministic commit order, the actual execution on different nodes might still diverge. This would lead to differences in this metric on different nodes and thus it would prevent nodes from reaching consensus.

To address this issue, we introduce an optimistic scheduler with deterministic transaction aborts. To our knowledge, this algorithm is the first of its kind, mostly because distributed ledgers have more stringent determinism requirements than most other domains. Based on our evaluation, this approach allows us to introduce incentives for parallelizability in exchange for a performance impact that is, on average, acceptable.

In summary, the major contributions of this paper are recognizing that certain common application-inherent transaction conflicts lead to bottlenecks under parallel execution, providing a set of effective techniques to deal with these, and offering a deterministic scheduling algorithm that makes it possible to incentivize better parallelism.

\section{Background and Motivation} \label{background}

Bitcoin \cite{nakamoto2008bitcoin} introduced \textit{blockchains} with the goal of supporting \textit{cryptocurrency} payment transactions without relying on any central authority.
Such a public blockchain is a distributed ledger maintained by a peer-to-peer network in a \textit{trustless} and \textit{permissionless} way.
The core piece of this technology is its \textit{consensus protocol}, \textit{Nakamoto consensus}, that probabilistically guarantees the irreversibility of transactions in decentralized public settings, even under adversarial conditions.
The ledger is composed of a chain of \textit{blocks}, each of which contains a sequence of transactions, and replicated among all the participant nodes.
Each block is generated by a \textit{miner} through some \textit{Proof-of-Work} mechanism, chained at the tail of the valid chain in the miner's view, and broadcast to all the other \textit{validator} nodes through a peer-to-peer gossip network.
Due to the latency of block propagation in the network, multiple miners may generate blocks concurrently without seeing the others, and hence may introduce \textit{forks} into the ledger.
The Nakamoto consensus employs the \textit{longest chain rule} to let all the honest nodes agree on the valid chain and execute the transactions according to the order of the blocks in the chain and the order of the transactions in each block.
The miner of each block on the valid chain gets a certain amount of bitcoin as a reward from the system.
The security guarantee is achieved when forks are rare and the ledger basically forms a single chain.
In order to avoid forks, the Bitcoin protocol dictates a very low block generation rate in the entire network, which seriously limits its throughput. Specifically, Bitcoin can only achieve a throughput of 7 transactions per second (tps).

Ethereum extends Bitcoin with support for a Turing-complete programming framework, and the Solidity programming language, which allows developers to implement complex \textit{decentralized applications}.
This makes it possible to apply blockchain in industries like financial systems, supply chains, and health care~\cite{DeloitteFinancial, IBMSupplyChain, DeloitteHealthCare}.
In Ethereum, the \textit{state} resulting from transaction execution is maintained in the form of a \textit{Merkle tree}.
Ethereum adopts an \textit{account} model in its state.
There are two types of accounts: \textit{user} accounts and \textit{smart contract} accounts.
A user account is associated with its \textit{ether} balance information while each smart contract account further has an associated executable code and its own storage represented as a collection of key-value pairs maintained in the Merkle tree.
Each transaction occurs between a \textit{sender} account and a \textit{recipient} account.
The majority of transactions are one of two kinds: either a \textit{value transfer}, which is a purely monetary transfer of \textit{ether} from sender to recipient, or a \textit{contract call}, where the sender account triggers execution of the code associated with the recipient account.
During its execution, a contract call transaction can call functions of other smart contracts.
To ensure that transaction execution terminates, each computational step incurs a cost denominated in \textit{gas}, paid by the transaction sender.
The sender specifies a maximum amount of gas it is willing to pay (\textit{gas limit}), and if the charge exceeds this value, the computation is terminated and rolled back, and the sender’s gas is not refunded.
The smart contract code consists of a sequence of bytecode instructions that can be interpreted and executed by the \textit{Ethereum Virtual Machine} (EVM) to manipulate the state of the Merkle tree by updating the values of the corresponding keys.
Every bytecode instruction consumes a certain amount of gas.
Smart contracts developed using Solidity are compiled into such bytecode sequence before they are published into the blockchain.

Like Bitcoin, Ethereum also employs Nakamoto consensus, although with some different system parameters, e.g., block size, block generation rate, etc. It improves the transaction throughput to about 30 tps but the consensus still remains the major performance bottleneck. In this situation, it makes sense that the EVM is designed as a single-thread engine without the need to introduce parallelism into the transaction execution.

To overcome the throughput bottleneck of Nakamoto consensus, many new and more advanced consensus protocols have been proposed in recent years~\cite{GHOST, BitcoinNG, Fruitchain, ohie, inclusive, PHANTOM, conflux, prism, Algorand}.
These protocols explore alternative structures to organize blocks, e.g., DAG-like structure, together with some novel deterministic block ordering schemes to allow faster global block generation rate without compromising the decentralization and security of the network, and hence the consensus mechanism ceases to be the system bottleneck.
For example, both Conflux~\cite{conflux} and OHIE~\cite{ohie} are able to process simple payment transactions with a throughput of more than 5000 tps, several orders of magnitudes faster than the original Nakamoto consensus.
Further research work like Shrec~\cite{shrec} also studies and develops a new transaction relay protocol that can more effectively utilize the network bandwidth to prevent it from becoming the new system bottleneck under high transaction throughput scenarios.
These techniques shift the throughput bottleneck of blockchain systems to the transaction and smart contract execution, therefore, introduce the pressing need for new technologies that can exploit the parallelism and increase the efficiency of transaction execution.

Some recent research works~\cite{saraph2019empirical, reijsbergen2020exploiting, anjana2019efficient, zhang2018enabling, pang2019concurrency, dickerson2019adding, doziercorrectness, bartoletti2020true} have explored the designs of a parallel smart contract virtual machine by integrating various mechanisms of concurrency control.
However, according to the reported results, the speedup that can be achieved by these proposed solutions is far from linear when applied to the real Ethereum workload.
We observed that this is mainly because of the lack of inherent parallelism in the real-world workload itself.
For example, by investigating the historical Ethereum workload, we found that many critical paths of a series of transactions that have to be executed sequentially are caused by the use of shared global counters.
We believe that the essential way to further improve significantly the inherent parallelism of the real workload is to introduce a better programming paradigm that can allow the developers to express parallelism more easily while keeping the original semantics.
In addition, in the decentralized environment, driving users to adopt a new paradigm is not that straightforward, as it may incur extra costs, from either the engineering or the economics considerations.
Therefore, some new design of incentive mechanisms is required to make the paradigm applicable to real-world applications.

\section{Empirical Study} \label{empirical_study}

What speedup should we expect when we execute blockchain transactions in parallel? To answer this question, we designed an empirical study using a dataset of historical Ethereum transactions.

\subsection{Methodology}

We empirically studied the amount of parallelism present in a real-world dataset using historical Ethereum transactions. To this day, Ethereum remains the backbone of the decentralized application ecosystem. As such, this workload represents the most common smart contract interaction scenarios, and the findings can be generalized to many other systems. Our experiment mainly focuses on the period between Jan-01-2018 and May-28-2018 ($858,236$ blocks in total), see Sections \ref{sec:generalizability} and \ref{threats_to_validity} for a more detailed justification of the dataset used.

The subject of this experiment is smart contract storage conflicts, i.e., cases where two transactions within the same block access the same entry in the state tree, and at least one of these accesses is a write. To obtain these results, we ran an OpenEthereum node (formerly Parity) modified so that it tracks and stores all contract storage accesses. We stored these traces for blocks \texttt{\#1} to \texttt{\#5692235} in a local database. In this experiment, other kinds of conflicting accesses (e.g., conflicts on the account balance) are not considered.

Given that the execution time of transactions is unknown and might vary from node to node, we used the transaction gas cost, obtained from the transaction receipt, as an approximation of this. This follows the practice of a number of related works \cite{saraph2019empirical, reijsbergen2020exploiting}.

Given the transaction dependencies derived from their state access traces and the gas costs of the transactions, we constructed a dependency graph for each block. Then, simulating non-preemptive execution on 2, 4, 8, 16, and 32 threads, we constructed an optimal schedule for each block, i.e., a schedule that ensures that no transaction needs to abort while also maximizing thread utilization. Under this execution model, the overall execution cost of this schedule puts an upper bound on the potential speedup that we can achieve; any other schedule might either need to abort and re-execute conflicting transactions, or delay execution through locking. Apart from the overall execution cost (as approximated through the overall gas cost), we also inspected the heaviest path in the transaction dependency graph.

\subsection{Results and Findings}
\label{sec:empirical-results}

\begin{figure}[h]
  \centering
  \includegraphics[width=\linewidth]{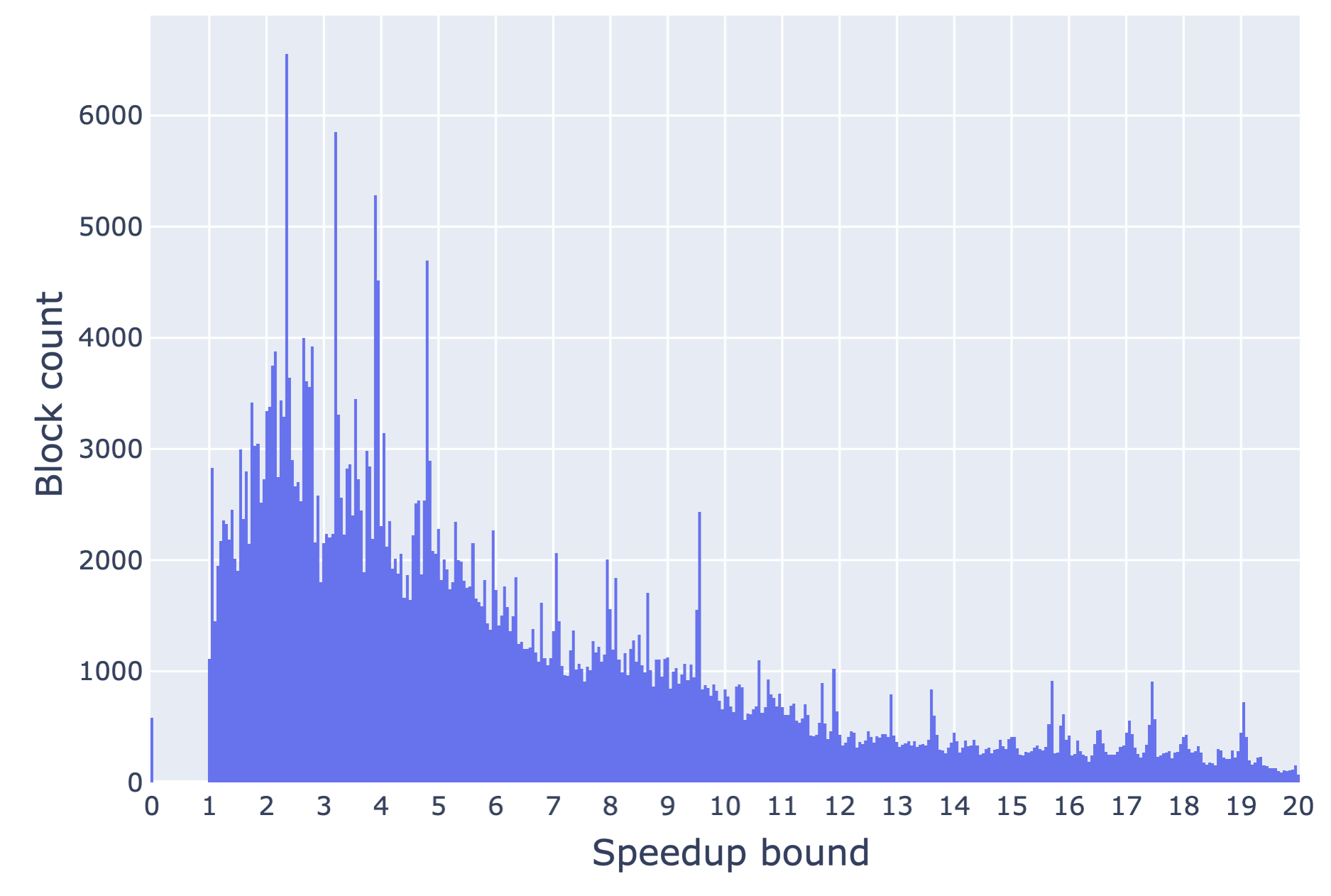}
  \caption{Distribution of parallel speedup bounds}
  \label{fig:speedup-bound}
\end{figure}

\paragraph{Execution Bottlenecks.}
The experiment shed some light on the limits of speedup we can expect to achieve when executing Ethereum transactions in parallel. We found that the overall speedup on the observed period was only 4x compared to the serial execution, an underwhelming result considering that we had 8, 16, or even more threads available. A closer look at the per-block results shows that in fact, many blocks have much higher speedups, but a significant portion of blocks perform poorly (see Figure \ref{fig:speedup-bound}).

When comparing the execution cost of a block to the execution cost of the heaviest path in its dependency graph, we found that these two often coincide. This means that the overall execution is bottlenecked on the execution of the heaviest path. When we look at single blocks, this heaviest path is often just a single transaction: When, for example, a block has many simple payment transactions and one expensive smart contract call that executes hundreds of token transfers, then this latter transaction will dominate the execution time.

Under our non-preemptive scheduler model and the inherently serial execution model of the EVM, there is no easy way to handle such single-transaction bottlenecks. Our focus, instead, is finding effective ways to handle bottleneck chains of two or more transactions. To focus on these, we re-ran our experiment with batches of consecutive blocks as the unit of execution, instead of just a single block. The idea is that, given thousands of transactions, the relative weight of a single transaction will be much smaller. The same experiment, executed on batches of 30 blocks, shows an overall speedup of 9.46x compared to serial execution. In this case, we observed the same result: Batches are often bottlenecked on a single chain of tens or sometimes hundreds of dependent transactions.

\begin{table*}
  \caption{Examples for bottleneck root causes from our 200-batch random sample}
  \label{tab:bottleneck-examples}
  \footnotesize
  \begin{tabular}{clclcl}
    \toprule
    \textbf{block batch} & \textbf{contract} & \textbf{contract type} & \textbf{method(s)} & \textbf{conflict type} & \textbf{conflict source} \\
    \midrule
    5536219-5536248                                            & Storj                        & ERC20         & \texttt{transfer} (STORJ)              & counter       & same sender account                 \\
    5559949-5559978                                            & Free BOB Tokens              & ERC20         & \texttt{airdrop} (BOBx)                & counter       & \texttt{totalSupply}                         \\
    5497669-5497698                                            & IDEX                         & DeFi          & \texttt{trade}, \texttt{adminWithdraw}          & counter       & ETH fee account balance             \\
    5493409-5493438                                            & Bancor                       & DeFi          & \texttt{quickConvert}                  & counter       & Bancor (BNT) fee recipient  \\
    5562289-5562318                                            & CryptoKitties: Core          & games/NFT     & \texttt{breedWithAuto}                 & counter       & \texttt{pregnantKitties++}                   \\
    5562409-5562438                                            & Mythereum Card               & games/NFT     & \texttt{mintSpecificCards}             & array         & \texttt{cards.push(card)}\\
    \bottomrule
  \end{tabular}
\end{table*}

We further examined the impact of these bottleneck transaction chains by re-running the experiment, while ignoring conflicts arising from these smart contracts. The result is an overall speedup of 23.8x compared to serial execution. These results show that bottleneck transactions not only have a crucial impact on the parallelism of our dataset, but also that by breaking up these dependency chains, we can potentially achieve significantly higher speedups.

\paragraph{Classification of Smart Contract Conflicts.}
To gain a better understanding of smart contract bottlenecks, we collected the primary bottleneck transaction chains for each 30-block batch, and collected the batches that have a speedup bound of 10x or less (3242 in total). Then, we selected a random sample of 200 batches and analyzed them manually. Table \ref{tab:bottleneck-examples} shows selected examples from this sample.

In terms of application types, we identified three broad categories: \textit{ERC20 tokens} (token distribution, airdrops) accounted for 60\% of the bottlenecks in our sample, \textit{Decentralized Finance} (DeFi) applications made up 29\%, while \textit{games and collectibles} (non-fungible tokens, NFTs~\cite{nfteip}) were the cause in 10\% of the cases.

In most cases, ERC20 tokens lead to conflicts when there are several token transfers over multiple transactions that distribute tokens from the same sender address.
Transactions might also have other dependencies, for instance, the total supply is updated every time new tokens are minted. While ERC20 token distributions are heterogeneous in their implementation (e.g., they use various interfaces like \texttt{transfer}, \texttt{multiTransfer}, \texttt{batchTransfer}, \texttt{multisend}, \texttt{aidrop}), these all result in similar conflict patterns.

In DeFi applications like IDEX and Bancor, a common source of conflict is the fee account whose token balance gets updated for every trade. In the case of IDEX, the majority of trades involve ETH, so they all increment the ETH balance of the IDEX fee account.

Examples for games and collectibles (NFTs) include CryptoKitties, Etheremon, and IdleEth. These often involve some globally shared counters, like the number of kitties in the case of CryptoKitties. Maintaining an array of game items is also common. When a game involves payments and rewards, the fee recipient and reward sender account's balance might also lead to storage conflicts.

In terms of the source of conflicts, we found that in 194 of 200 batches (97\%) the root cause is one or more counters that get incremented (or decremented) by different transactions. In our sample, the other common source of conflicts, arrays, only accounted for about 2\% of the cases.

\paragraph{Bottleneck Code Examples.}
As an example for counter conflicts in token distributions, let us discuss the example in Listing ~\ref{lst:sol-counter}. When calling \texttt{transfer}, the sender's balance (\texttt{balances[msg.sender]}) is debited, while the recipient's balance is credited. The sender's balance corresponds to one specific storage location in the state tree. The debit operation will compile to a load (\texttt{SLOAD}), an add (\texttt{ADD}), and a store (\texttt{SSTORE}) operation, among others. When two transactions trigger this function from the same sender address concurrently, this will result in a conflict.

\begin{lstlisting}[language=Solidity, label={lst:sol-counter}, caption={Solidity counters (source: ConsenSys EIP20.sol)}]
function transfer(address _to, uint256 _val) /* ... */ {
    balances[msg.sender] -= _val; // <<<
    balances[_to] += _val;
    // ...
}
\end{lstlisting}

Let us look at another example, this time for arrays and collectibles (Listing ~\ref{lst:sol-array}). In the popular CryptoKitties Ethereum game, each new collectible is stored in an array. The \texttt{push} operation on Solidity arrays will modify two storage entries: First, it will store the new item at a location derived from the array's length, and second, it will increment its length. Two concurrent transactions will both modify the array length and as such, they will conflict.

\begin{lstlisting}[language=Solidity, label={lst:sol-array}, caption={Solidity arrays (source: CryptoKitties)}]
function _createKitty(/* ... */) /* ... */ {
  uint256 newKittenId = kitties.push(_kitty) - 1; // <<<
  // ...
}
\end{lstlisting}

\subsection{Generalizability of the Observations}\label{sec:generalizability}

Our evaluations are based on a relatively narrow period of the Ethereum transaction history.
This is because acquiring the entire transaction dataset and generating storage access traces is extremely resource-consuming, both in terms of storage and time.

We believe the chosen period is representative of today's Ethereum workload and so our findings are generalizable.
The application patterns we observed (DeFi, NFT, token distributions) are even more dominant today.
Contract developers have no incentive to address common storage bottlenecks.
In fact, just by a cursory glance, we can spot storage conflicts in many recent popular applications:
Uniswap exchanges that involve the same token will always conflict on the counters that represent token reserves (contract \texttt{UniswapV2Pair}).
Similarly, OpenSea trades will transfer tokens to the same \texttt{protocolFeeRecipient}.
This suggests that the conflicts we identified are even more common today.
\section{Avoiding Application Inherent Conflicts} \label{avoid_conflicts}

As we have seen in Section \ref{empirical_study}, a large portion of storage conflicts is associated with storage slots that belong to either counters or arrays. By \textit{counter} here we mean a variable that one can use to track a quantity by incrementing or decrementing it, regardless of its current value. \textit{Arrays} in Solidity are a simple data structure that stores a sequence of elements, along with the number of elements.

In theory, a transaction dependency chain could involve multiple conflicting storage slots. For instance, the chain \texttt{\#a <-- \#b <-- \#c} could mean that \texttt{\#a} and \texttt{\#b} conflict on a counter, while \texttt{\#b} and \texttt{\#c} conflict on an unrelated array. In practice, however, this is rarely the case. Most transactions in a conflict chain will execute similar operations and will conflict on the same storage entry or entries. In this case, dependencies are \textit{transitive}, i.e., \texttt{\#c} will conflict with \texttt{\#a}.

\begin{figure}[h]
    \centering
    \includegraphics[width=0.45\textwidth]{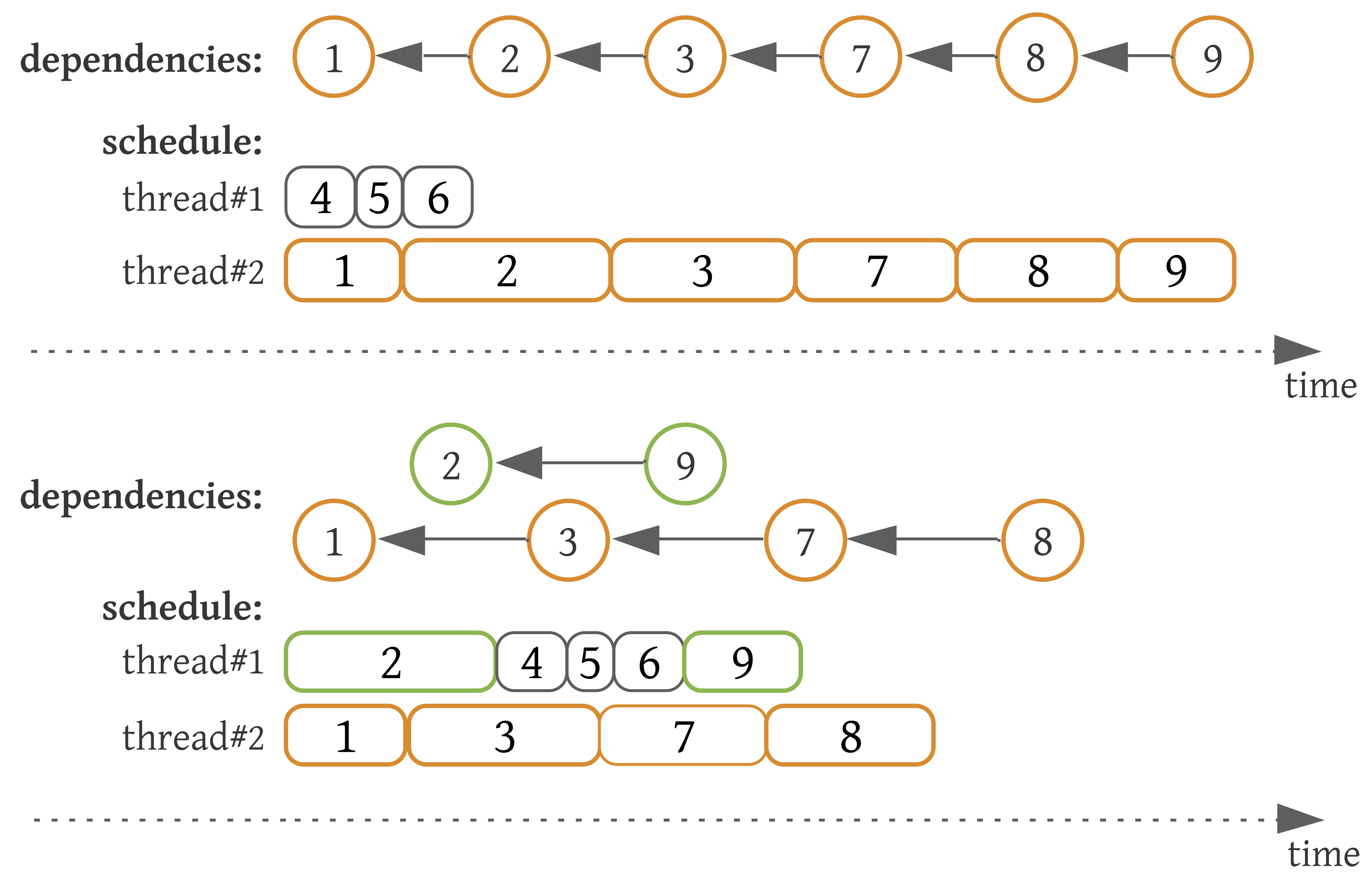} %
    \caption{Breaking up a conflict chain into multiple disjoint conflict chains. On the top of the figure, a long conflict chain requires transactions \texttt{\#1}-\texttt{\#2}-\texttt{\#3}-\texttt{\#7}-\texttt{\#8}-\texttt{\#9} to be scheduled serially on the same thread, dominating the overall execution time. By breaking up this chain into two (\texttt{\#1}-\texttt{\#3}-\texttt{\#7}-\texttt{\#8} and \texttt{\#2}-\texttt{\#9}), each resulting chain will still need to be executed serially, but the two chains can be executed in parallel to each other. This allows us to achieve a much higher speedup.}
    \label{fig:breaking-up-chain}
    \vspace{-3ex}
\end{figure}

To alleviate the impact of these transaction bottleneck chains, we need to \textit{break them up} into multiple shorter chains by eliminating dependencies between subsets of the transactions involved (see Figure \ref{fig:breaking-up-chain}). We propose three techniques to achieve this.
As arrays only account for a small fraction of storage bottlenecks (Section \ref{sec:empirical-results}), we will focus on counters in this section.

\paragraph{\textbf{Technique 1: Conflict-Aware Token Distribution.}} In our evaluations, we saw that token distributions (token sales, airdrops) are by far the most common sources of bottleneck conflicts. In the majority of cases, the source of conflict is the storage entry that stores the sender account's current balance.

The simplest way to address these common bottlenecks is to use multiple sender addresses. By distributing the initial funds (where applicable) to a set of sender accounts instead of a single account, and using different sender addresses for consecutive transactions, we can divide the set of bottleneck transactions into disjoint sets of conflicting transactions, each less likely to form a bottleneck.

Of course, the feasibility of this approach depends on the specific implementation of the token. Some tokens have other dependencies: for instance, the total supply of tokens might also be incremented each time new tokens are minted. In the presence of such dependencies, we need a more sophisticated and general approach.

\paragraph{\textbf{Technique 2: Partitioned Counters.}} Using a technique similar to \textit{sloppy counters} widely used in the Linux kernel \cite{boyd2010analysis}, we propose a way to route multiple writes on the same counter to multiple distinct storage entries. As writes to different storage entries do not conflict, this technique can drastically reduce the conflict rate.

The main idea of partitioned counters is shown in Listing \ref{lst:sol-partitioned}. Here we have a single contract that represents a counter instance. The value of the counter is actually maintained on 3 separate storage entries called \textit{sub-counters}. Each time a transaction modifies the counter's value, we assign a sub-counter based on the transaction's sender address. As addresses are derived using cryptographic hashing, this can be viewed as a pseudorandom sub-counter assignment. When reading the value of the counter, all sub-counters are accessed and their values are summed.

\begin{lstlisting}[language=Solidity, label={lst:sol-partitioned}, caption={Partitioned counters implemented in Solidity}]
contract PartitionedCounter { // LEN = 3
  int256[LEN] public cnt;

  function add(uint32 n) internal {
    uint8 slot = uint8(tx.origin) % LEN
    cnt[slot] += n;
  }

  function get() internal view returns (int256 sum) {
    for (uint8 i = 0; i < LEN; ++i) { sum += cnt[i]; }
  }
}
\end{lstlisting}

Partitioned counters have several advantages. First, a given transaction's writes will all operate on a single storage entry, even if it increments the counter multiple times, as the sender address does not change throughout the transaction's execution. Second, two transactions from two distinct sender addresses that both increment the counter have a much-reduced chance of operating on the same sub-counter and thus conflicts are often avoided. Third, the counter can be adjusted based on the use case, e.g., for counters used frequently one could use more sub-counters, and one could use different criteria for routing transactions to different sub-counters. Our example routes transactions based on the sender address (\texttt{tx.origin}) as this addresses common token conflicts.

Partitioned counters have two main drawbacks. First, while we only need to access a single storage entry for writing the counter, reading it will touch all sub-counters. As a result, any transaction that reads the counter will conflict with all writing transactions. As such, this technique is suitable for write-heavy counters. Fortunately, many of the counters we analyzed are never read through transactions. Second, partitioned counters can be significantly more expensive than built-in integers, especially when it comes to reading the counter. This drawback is offset by the potential increase in parallel speedup that partitioned counters offer. Moreover, many counters are rarely or never read in a transaction context.

\paragraph{\textbf{Technique 3: Commutative EVM Instructions.}} We have discussed two approaches. One operates on the \textit{application level}, i.e., it addresses conflicts by introducing specific ways to interact with the application. The other operates on the \textit{smart contract level}, by offering tools to contract developers to avoid conflicts. A third approach is to tackle conflicts on the \textit{virtual machine level} by extending the protocol by new instructions that have better conflict tolerance.

When the \textit{Ethereum Virtual Machine (EVM)} executes an increment operation, it first loads the storage entry's current value into memory (\texttt{SLOAD}), then modifies this value (\texttt{ADD}), and finally it stores the end result back into the storage entry (\texttt{SSTORE}). This behavior originates from the Solidity compiler. As discussed before, two transactions incrementing the same counter will both read and write the same storage entry, and so they will conflict.

For counter increments, the current value is only used for calculating the new value, and otherwise it is irrelevant. Put in another way, unlike other read-write conflicts, increments are \textit{commutative}. Two transactions that increment the same counter, but do not use its value otherwise, could be executed in any order. However, under the current semantics of the EVM, such transactions will conflict.

We introduce special semantics for executing increments in a way that does not result in conflicts. Rather than compiling increments into \texttt{SSLOAD} and \texttt{SSTORE} instructions, they instead get compiled into a single \texttt{CADD} instruction that stands for \textit{commutative add}. This instruction takes a storage location and a value as its parameters. When the VM encounters a \texttt{CADD} instruction, it does not eagerly execute the addition, but instead, it records this operation in an in-memory temporary storage.
If the VM encounters an \texttt{SSTORE} operation, it then erases the pending \texttt{CADD} instructions on the same storage location as they have been overwritten.
If the VM encounters an \texttt{SLOAD} operation, it then first executes all pending \texttt{CADD} operations on the same storage location, then uses the result for this \texttt{SLOAD}.

After the transaction has been executed, the scheduler proceeds to check for conflicts. Concurrent storage reads and writes to the same storage location constitute conflicts. If, however, two transactions only have \texttt{CADD} operations on a storage location, but no other reads, then they are not considered conflicting. In this case, these \texttt{CADD} operations are executed serially during the commit phase.

Introducing a \texttt{CADD} instruction for signaling commutative operations to the VM allows us to avoid a major class of transaction conflicts that originate from operations on a single counter.

\section{OCC With Deterministic Aborts} \label{deterministic_scheduling}

\subsection{Incentives in Parallel Scheduling}
\label{sec:incentives}

Permissionless blockchains have no central authority that could enforce protocol compliance. Instead, protocol designers introduce \textit{incentives} that encourage \emph{good} behavior (creating blocks, avoiding storage bloat) and penalize \emph{bad} behavior (attacks). The efficiency of parallel schedulers depends on various factors, some of which are under the users' control. Therefore, parallel execution must also come with a set of incentives that maximize its effectiveness.

A detailed design of such a system of incentives is beyond the scope of this paper. We observe, however, that any incentive system must be able to deal with spam or Denial of Service attacks that target mispriced operations and resources in the system, as has happened several times on Ethereum \cite{chen2017adaptive}. Parallel execution based on OCC will inevitably lead to some transaction aborts and re-executions. If there is a way for users to intentionally trigger aborts without any penalty, then that opens up the door to a serious DoS vulnerability of the scheduler. Our goal, then, is to define an execution framework that would allow schedulers to deal with this issue by deterministically pricing transaction re-executions.

\begin{figure*}
    \centering
    \begin{minipage}[t]{0.33\textwidth}
        \centering
        \captionsetup{width=0.93\linewidth}
        \includegraphics[width=0.95\textwidth]{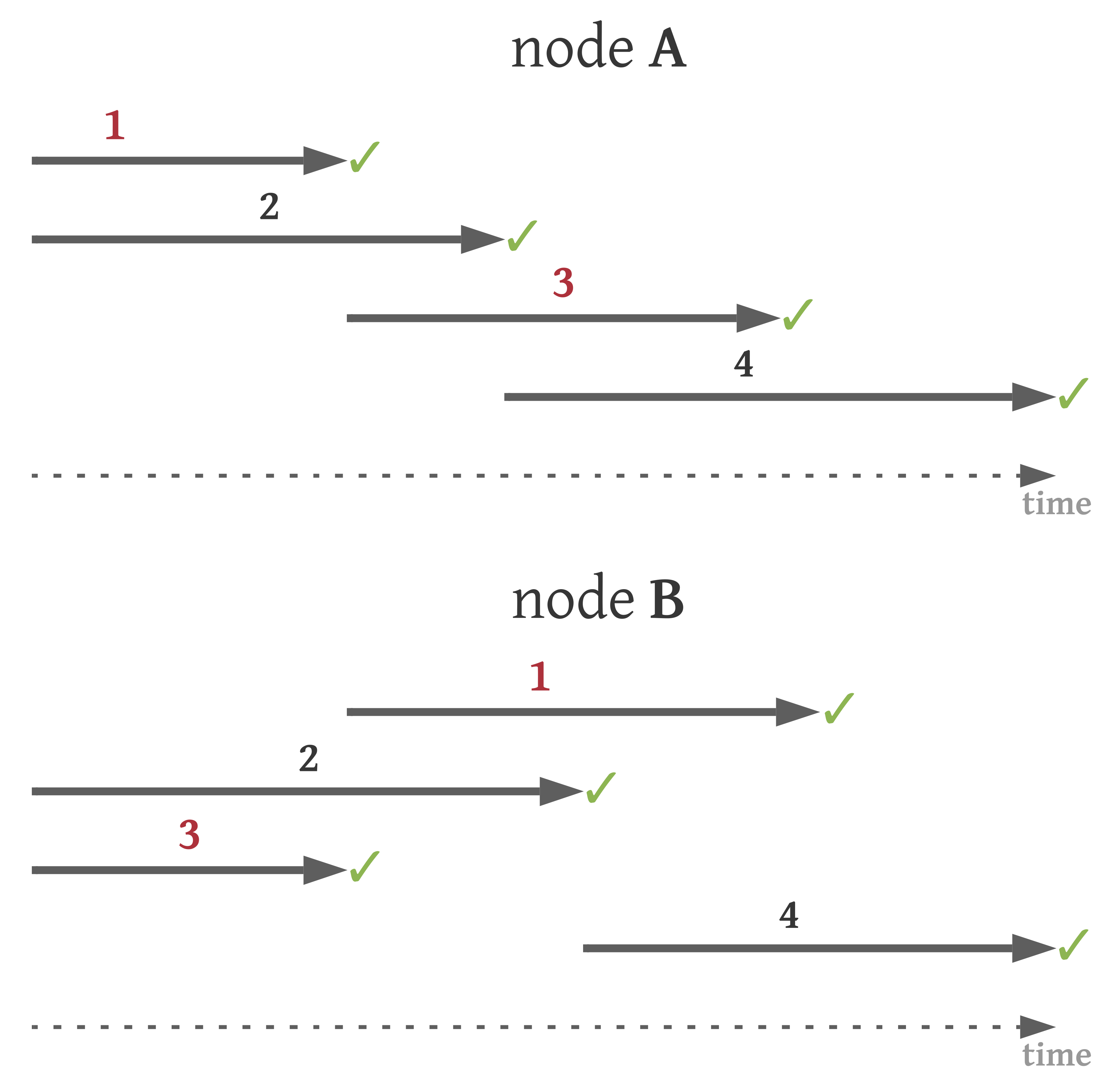} %
        \caption{Classic OCC: Transactions are committed right after execution, regardless of their order in the block. This results in different commit orders (\#1-\#2-\#3-\#4 and \#3-\#2-\#1-\#4) and end states might diverge.}
        \label{fig:example-occ-classic}
    \end{minipage}\hfill
    \begin{minipage}[t]{0.33\textwidth}
        \centering
        \captionsetup{width=0.93\linewidth}
        \includegraphics[width=0.95\textwidth]{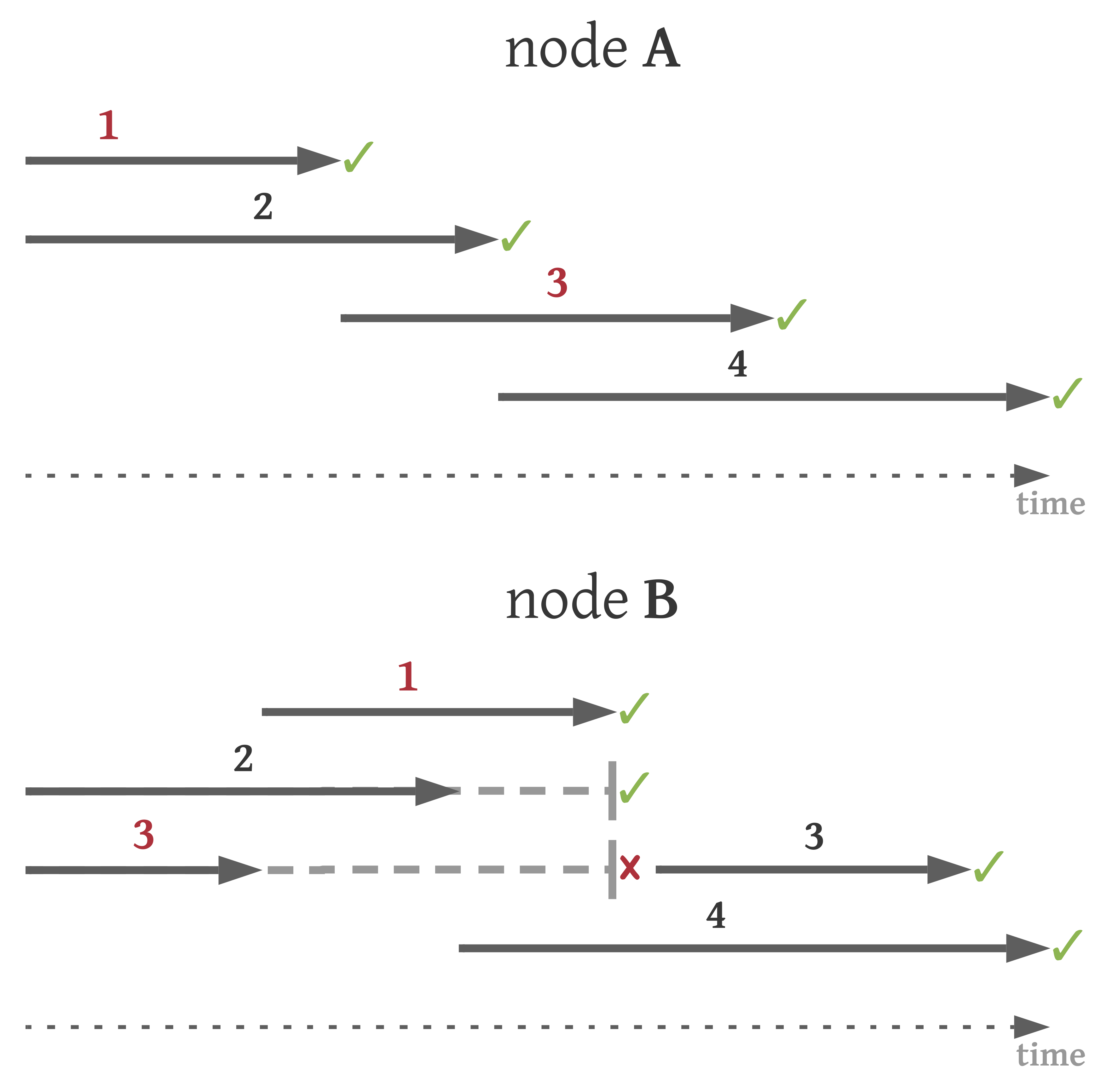} %
        \caption{OCC with det. commit order: After execution, commit is delayed until the previous transaction in the block has committed. The commit/abort decision for a transaction might diverge on different nodes (\#3).}
        \label{fig:example-occ-det-commit}
    \end{minipage}
    \hfill
    \begin{minipage}[t]{0.33\textwidth}
        \centering
        \captionsetup{width=0.93\linewidth}
        \includegraphics[width=0.95\textwidth]{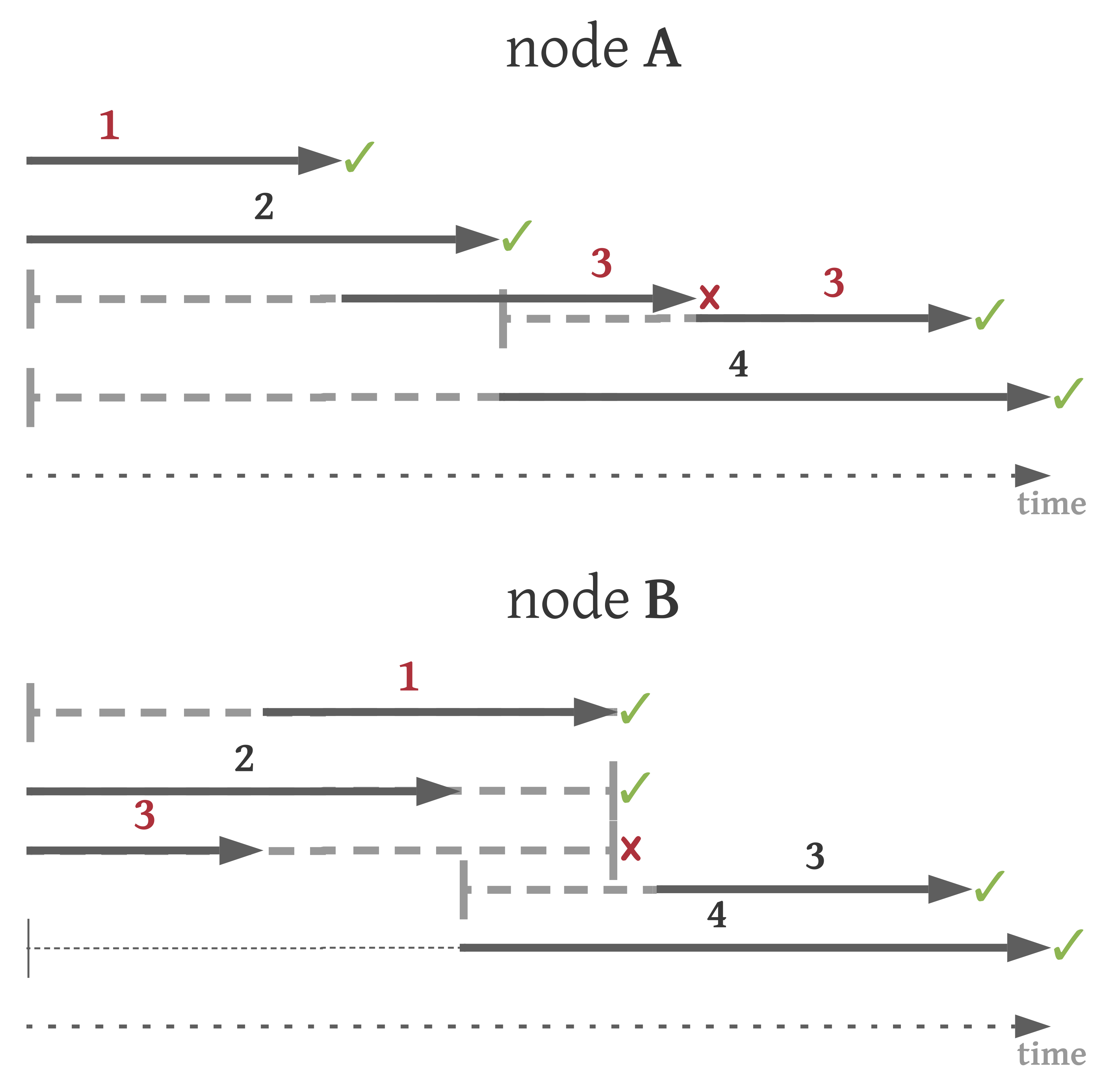} %
        \caption{OCC-DA: Transactions can only see a version of the state decided prior to execution, even if a more recent version is available. Each execution of a transaction will either commit or abort on all nodes.}
        \label{fig:example-occ-det-aborts}
    \end{minipage}
\end{figure*}

\subsection{Levels of Determinism}
\label{sec:levelsofdet}

Parallel schedulers introduce a level of non-determinism into the execution, as the precise timing of transactions might differ from node to node. This is in direct conflict with the requirements of the consensus mechanism, which relies on strict determinism for the nodes to converge into a consensus state. In blockchain systems, therefore, parallel schedulers must maintain higher levels of determinism compared to traditional algorithms.
We define the following three levels of determinism in optimistic transaction execution.

\begin{enumerate}

    \item \textbf{Classic OCC}: Classic OCC \cite{kung1981optimistic} has no determinism guarantees. Generally, transactions start execution on a first-come-first-served basis. Node-local consistency is typically ensured by the property of \textit{serializability}, which dictates that the observable results of the parallel execution are equivalent to those of \textit{some} serial execution.
    However, execution of the same transaction set on different nodes might correspond to different serial executions and yield diverging results.

    \item \textbf{OCC with deterministic commit order}: Instead of dictating that the parallel schedule is equivalent to \textit{any} serial schedule, it must correspond to a \textit{specific} serial schedule. This means that the final execution result on different nodes will be equivalent, even though the actual execution might differ. This requirement can be satisfied by committing transactions strictly according to the block transaction order, or by scheduling according to a dependency graph~\cite{anjana2019efficient}.

    \item \textbf{OCC with deterministic aborts}: While deterministic serialization order guarantees that the observable outputs (the resulting state) are the same across all nodes, the actual execution might still differ: Due to different timing of transactions, a transaction might be committed on one node, and aborted on another. If the protocol relies on this commit/abort decision to penalize aborts and avoid DoS attacks (see Section \ref{sec:incentives}), this will lead to diverging states. Thus, the highest level of determinism we aim for is when aborts themselves are deterministic: if a transaction is aborted once on one node, it is aborted exactly once on all the other nodes as well.
\end{enumerate}

\textit{OCC with deterministic commit order} is a topic with considerable research attention in deterministic database systems~\cite{thomson2010case, thomson2011building, thomson2012calvin, abadi2018overview}. On the other hand, the stringent requirements of \textit{OCC with deterministic aborts}, to the best of our knowledge, have not been described elsewhere. While imposing such restrictions on OCC schedulers might certainly have a negative impact on the parallel speedup, we argue it is crucial for implementing parallel schedulers under a distributed consensus setting.

\subsection{OCC-DA: OCC with Deterministic Aborts}

Our execution model is based on OCC with \textit{snapshot isolation}. Transactions are scheduled on a set of threads for execution. Executed transactions are committed according to the block transaction order. At the start of its execution, each transaction receives a \textit{snapshot} corresponding to the version of the storage after some transactions preceding it have been committed. This snapshot does not change during the execution of the transaction. The highest transaction id whose committed writes are part of this snapshot corresponds to the \textit{storage version} of the snapshot, or, equivalently, the storage version of the transaction to-be-scheduled.

As an example, let us assume that transaction \texttt{\#1} has been committed, transaction \texttt{\#2} is being executed on one thread, and we are scheduling transaction \texttt{\#3} on another thread. In this case, \texttt{\#3} can see storage version \texttt{\#1} (i.e., the contents of storage up to and including \texttt{\#1}'s writes). If, during the execution of \texttt{\#3}, \texttt{\#2} modifies some storage values, these updates are not visible to \texttt{\#3}. If, during the commit of \texttt{\#3}, the scheduler detects that some values read by \texttt{\#3} were concurrently modified by \texttt{\#2} and thus \texttt{\#3} operated on outdated values, then \texttt{\#3} is aborted and scheduled for re-execution.

In distributed consensus, transaction execution is deterministic: The same code triggered with the same inputs (its parameters and the current state) will produce the same outputs. From this, it is easy to see that a transaction executed over a specific storage version (i.e., the same state) on two different nodes will either commit on both or abort on both.

We then define OCC-DA as follows. We assign a storage version to each execution of each transaction \textit{prior to execution}: $(tx_{n}, i) \rightarrow sv_{n, i}$.
$(tx_n, i)$ stands for the $i$'s execution of transaction \#n, where $i = 0, 1, 2, ...$. Note that, depending on the scheduler implementation, a transaction can be executed two or more times. The last execution must commit, while all preceding executions will be aborted. For all potential executions $i$ of all transactions \#n in an execution unit (e.g., in a block), $sv_{n, i}$ is defined uniformly on all nodes, and it is defined prior to execution so that it does not rely on non-deterministic execution details. Then, for any $(tx_n, i)$, transaction \#n will either abort or commit on all nodes.

Throughout the execution of $(tx_n, i)$, the scheduler must allow the transaction to access storage entries written by transactions up to and including transaction $sv_{n, i}$. The scheduler must not allow the transaction to access storage entries written by a transaction with an id higher than $sv_{n, i}$, even if it is committed. If $sv_{n, i}$ has not committed and therefore the storage version specified prior to execution is not available when $(tx_n, i)$ is being scheduled for execution, the transaction cannot start execution and must wait. 

\subsection{Example}

We have 4 transactions, labeled \texttt{\#1}-\texttt{\#4}. Transactions \texttt{\#1} and \texttt{\#3} have a storage conflict: \texttt{\#1} writes a storage entry read by \texttt{\#3}. Let us then walk through scheduling these four transactions on two different nodes with 2 threads each, under different determinism guarantees.

Figure \ref{fig:example-occ-classic} depicts an example schedule using \textit{classic OCC}. This approach has no determinism guarantees. In particular, we can see that the commit order on node \textit{A} is \texttt{\#1}-\texttt{\#2}-\texttt{\#3}-\texttt{\#4}, while it is \texttt{\#3}-\texttt{\#2}-\texttt{\#1}-\texttt{\#4} on node \textit{B}. The diverging relative order of the two conflicting transactions (\texttt{\#1}-\texttt{\#3}, \texttt{\#3}-\texttt{\#1}) might lead to diverging states on the two nodes. While \texttt{\#1} and \texttt{\#3} conflict, in this example they are not executed concurrently and therefore neither needs to be aborted.

In Figure \ref{fig:example-occ-det-commit}, we see an example of \textit{OCC with deterministic commit order}. On node \textit{B}, \texttt{\#3} finishes execution before \texttt{\#1}. However, it is is not committed until after \texttt{\#1} has, at which point the conflict is detected and \texttt{\#3} is aborted. The final commit order on both nodes \textit{A} and \textit{B} is \texttt{\#1}-\texttt{\#2}-\texttt{\#3}-\texttt{\#4}. However, due to the different relative order of the execution of \texttt{\#1} and \texttt{\#3} on the two nodes, the first execution of \texttt{\#3} commits on node \textit{A} while it aborts on node \textit{B}.
In distributed consensus, such non-determinism is not acceptable (Section \ref{sec:levelsofdet}).

Note that \texttt{\#4} on node \textit{B} cannot read \textit{uncommitted} results from \texttt{\#2} or \texttt{\#3}, even though both finish execution before \texttt{\#4}. This kind of \textit{snapshot isolation} allows us to avoid \textit{cascading aborts}. An investigation of whether allowing transactions to read uncommitted results is beneficial is beyond the scope of this paper.

Finally, Figure \ref{fig:example-occ-det-aborts} shows how OCC-DA works. Prior to execution, all nodes decide that the first execution of \texttt{\#3} can only read the state prior to \texttt{\#1}'s execution ($sv_{3, 0} := 0$), while the second execution can see the state after \texttt{\#2} ($sv_{3, 1} := 2$). (The rationale for these values is discussed in Section \ref{seq:assigning_sv}.)
As a result, even though \texttt{\#3} is scheduled after \texttt{\#1} on node \textit{A}, it is not allowed to see \texttt{\#1}'s writes and thus it will abort. This yields a result consistent with the other case where \texttt{\#3} is executed concurrently with \texttt{\#1}, as on node \textit{B}. The second execution will see the latest state on both nodes \textit{A} and \textit{B} and consequently it will commit on both nodes.

\subsection{Assigning Storage Versions}\label{seq:assigning_sv}

Let us make some remarks about the assignment of storage versions. The simplest approach is to set $sv_{n, 0} := -1$. This approach does not rely on any information about the transaction set.
While this simple first approximation works, setting $sv_{n, 0}$ to $-1$ (the state prior to transaction \#0's execution) will lead to aborts if the transaction set contains any dependencies.

For a more sophisticated heuristic for storage version assignment, we can rely on two kinds of information.
First, we can use the expected execution time of transactions to find the latest storage version a transaction is expected to see. If, based on this estimation, \texttt{\#3} will start execution after \texttt{\#1} but before \texttt{\#2}, then we set $sv_{3, 0} := 1$.
Second, an estimation of the transaction dependency graph might allow us to prevent aborts. For instance, if we guess that \texttt{\#3} is likely to conflict with \texttt{\#1}, then we can set $sv_{3, 0} >= 1$.
We do not have perfect information about execution times or transaction dependencies. For the former, the transaction \textit{gas limit} can serve as a reasonable first estimation. For the latter, static analysis and various heuristics might provide us with an approximate dependency graph.

The accuracy of the storage version assignment has a direct effect on the performance of the parallel scheduler: If we use a storage version that is too low, then we risk introducing more aborts. If, on the other hand, we use a storage version that is too high, then the transaction might need to be delayed while it waits for the storage version to become available, leading to thread under-utilization.

Finally, another aspect to consider is the overhead of the scheduler. Maintaining multiple storage versions might introduce a significant storage overhead in case there are many writes. Limiting the lowest storage version each transaction can see might help us put a limit on this overhead.

\subsection{The Algorithm}
A detailed algorithm for OCC-DA is presented in Algorithm \ref{listing:detocc}.
This algorithm takes a set of transactions and their dependencies as inputs. 
The dependency graph can be constructed through an estimation of the read-write set of each transaction. 
It is not necessary for the estimation to be perfect but it needs to be deterministic and consistent on all the blockchain nodes. 
The more precise it is, the fewer unnecessary aborts we may encounter.

\begin{algorithm}
\small
\caption{OCC-DA}
\label{listing:detocc}
\KwIn{Transactions $T$, gas $Gas$, number of threads $t$, none or a dependency graph $D$}
$H_{txs} \gets$ an empty minheap of $(sv, id)$ \;
\For{$id \in [0, |T|)$}{
    \eIf{$D$ exists}{
        $id_{max} \gets -1$ \;
        \For{edge $(id, id_{prev}) \in D$} {
            \Comment{tx\_$id$ depends on tx\_$id_{prev}$}
            \Comment{tx\_$id$ reads what tx\_$id_{prev}$ writes}
            $id_{max} \gets max(id_{max}, id_{prev})$ \;
        }
        $H_{txs}.push((id_{max}, id))$ \;
    }{
        $H_{txs}.push((-1, id))$ \;
    }
}
$H_{ready} \gets$ an empty minheap of $(id, sv)$ \;
$H_{threads} \gets$ an empty minheap of $(gas, id, sv)$ \;
$H_{commit} \gets$ an empty minheap of $(id, sv)$ \;
$next \gets 0$ \;
\While{$next < |T|$}{
    \Comment{Stage 1 : Schedule}
    \For{$(sv, id) \gets H_{txs}.pop()$}{
        \eIf{$sv > next - 1$}{
            $H_{txs}.push((sv, id))$ \;
            \Break
        }{
            $H_{ready}.push((id, sv))$ \;
        }
    }
    \While{$|H_{threads}| < pool\_size$ and $|H_{ready}| > 0$}{
        $(id, sv) \gets H_{ready}.pop()$ \;
        $H_{threads}.push((Gas[id], id, sv))$ \;
    }
    \Comment{Stage 2 : Execution}
    \If{$|H_{threads}| > 0$}{
        $(gas, id, sv) \gets H_{threads}.pop()$ \;
        $H_{commit} \gets (id, sv)$ \;
        \For{$i \in [0, |H_{threads}|)$}{
            $H_{threads}[i].gas \gets H_{threads}[i].gas - gas$ \;
        }
    }
    \Comment{Stage 3 : Commit/Abort}
    \While{$|H_{commit}| > 0$}{
        $(id, sv) \gets H_{commit}.pop()$ \;
        \If{$id \neq next$}{
            $H_{commit}.push((id, sv))$ \;
            \Break
        }
        \For{$id_{prev} \in [sv + 1, id - 1]$}{
            \If{tx\_$id_{prev}$'s write set $\cap$ tx\_$id$'s read set $\neq \emptyset$}{
                get Aborted \;
                \Break
            }
        }
        \eIf{Aborted}{
            $H_{txs}.push((id - 1, id))$ \;
        }{
            \Comment{Commit successfully}
            $next \gets next + 1$ \;
        }
    }
}
\Return{}
\end{algorithm}

In the beginning, the storage version of each transaction is initialized as the maximum id of the transactions that it depends on according to the dependency graph, or $-1$ if it has no dependency (lines 2-11). 
The transactions are pushed into a min-heap $H_{txs}$ indexed by the storage version. 
There are three other min-heaps. 
$H_{ready}$ maintains transactions ready to be scheduled, $H_{threads}$ is exactly the thread pool for executing transactions, and $H_{commit}$ is for the transactions that have already finished the execution and wait to be committed.
The global variable $next$ maintains the id of the next to-be-committed transaction.
Note that the algorithm describes the transaction execution mechanism used in our simulation which results in deterministic execution completion order according to the given gas consumption of each transaction. 
However, in a real system, the correctness and effectiveness of our scheduling strategy do not rely on this execution determinism.

Lines 16-47 show the stages that transactions experience. Stage 1 is scheduling transactions into the thread pool (17-26). We consider a transaction ready to execute when the transaction corresponding to its storage version has been committed. 
Ready transactions are pushed into the thread pool, if it has empty slots (24-26).

Stage 2 is the execution of transactions in the thread pool. We simply choose the transaction with the minimal remaining gas, which is exactly the top of $H_{threads}$, push it into $H_{commit}$, and maintain the gas accordingly.

The last stage is trying to commit the transactions one by one in $H_{commit}$ (lines 33-47). Transactions in $H_{commit}$ are maintained in the order of id, since we always commit transactions in order without skips. 
For each to-be-committed transaction, the algorithm checks whether it should be aborted or committed through checking whether there exist any read-write conflicts between the current transaction and those transactions committed since it starts to execute (lines 39-42). If aborted, the transaction is pushed back into $H_{txs}$ with its new storage version set as $id - 1$, otherwise, the commit succeeds.
\section{Evaluation} \label{evaluation}

\subsection{Experimental Setup}

The experimental evaluation of the techniques presented in this paper builds on the empirical study discussed previously. All simulations discussed here operate on the storage access traces collected from the Ethereum transaction dataset, as outlined in Section \ref{empirical_study}.

For evaluating the proposed bottleneck-elimination techniques, we analyzed the best-case parallel execution time using the transaction dependency graph, with and without applying these techniques. 
In this experiment, we rely on perfect knowledge of transaction dependencies. We start by constructing a dependency graph of transactions, where vertices (that correspond to the transactions) are weighted by the transaction gas costs. Then, we simulate scheduling the transactions on a set of threads (2, 4, 8, 16, 32). In each scheduling step, out of all transactions with no unexecuted dependencies, we select the one that has the heaviest path starting from it. The gas cost of this schedule serves as the baseline. For this experiment, we use 10-block batches as the unit of execution, to reduce the effect of single-transaction bottlenecks (see Section \ref{sec:empirical-results}).
For evaluating the potential effect of using partitioned counters, we prune the transaction dependencies in a pseudorandom fashion, in a way that is consistent with this technique. For instance, for a counter of length 3, for each dependency, we remove it with a probability of 8/9.

For seeing the impact of deterministic scheduling, we simulated an OCC scheduler with deterministic commit order as our baseline.
The scheduler has zero upfront information about the transaction dependencies.
When scheduling a transaction for execution, the scheduler uses the highest committed transaction id as its storage version.
To make the transaction commit order deterministic, the scheduler commits transactions according to their block order.
For deterministic aborts, instead of using the highest executed transaction id as the transaction's storage version, we use a storage version defined prior to execution: We use -1 (i.e., the storage prior to the block's execution) as the storage version for the transaction's first execution, and use $(tx\_id - 1)$ for its second execution. We then compare the overall gas costs of these two OCC simulations.

\subsection{Assumptions and Limitations}

Our evaluations are based on simulations using real-world data, not on implementation in a real-world blockchain system. We chose to simulate scheduling because currently there is no parallel EVM available. The potential overhead of the parallel VM is not considered as it depends heavily on the actual implementation. We aim to compare different techniques in a controlled and deterministic way without speculating about the VM implementation. For the two experiments, our focus is on the change in the maximum speedup achievable and the difference between the two OCC methods. We believe these relative measures also apply in a real-world system.

The evaluation assumes that partitioned counters can be applied to all storage conflicts. This is a reasonable approximation based on the results presented in Section \ref{sec:empirical-results}, where we found that almost all conflicts can be traced back to counters used in token distribution scenarios. The information available to our simulated scheduler (storage read/write traces) is insufficient to decide whether a storage location corresponds to a counter; for this, one would need to rely on the contract's source code, which is often not publicly available.

The other two proposed techniques are not evaluated.
The effect of \emph{"Conflict-Aware Token Distribution"} is equivalent to that of partitioned counters.
As for commutative EVM instructions, our storage traces do not provide information about the series of instructions executed that would be necessary to evaluate commutative operations.
We expect the effect of the three techniques to be similar, as they all remove edges from the transaction dependency graph in a similar way.
In fact, \emph{"Commutative EVM Instructions"} might be more effective:
The first two techniques will still result in some conflicts, while this third one could serialize all updates without conflict.

\subsection{Evaluation Results}

\paragraph{Overall Results.} For each 10-block batch, we look at its \textit{optimal} execution cost on 32 threads (based on the transaction dependency graph), and compare this to its serial execution cost. For the baseline (with no modification), the average speedup over all batches is 11.93x, while the overall speedup on the whole period is 9.25x, due to the bottlenecks discussed in Section \ref{empirical_study}. Using a counter of length 2, the average speedup becomes 21.23x, while the overall speedup is 17.96x. The highest speedup we can hope to achieve, when we remove all transaction dependencies, is 23.63x on average, while the overall speedup is 20.61x in this experiment.

\begin{figure}
    \centering
    \includegraphics[width=0.3\textwidth]{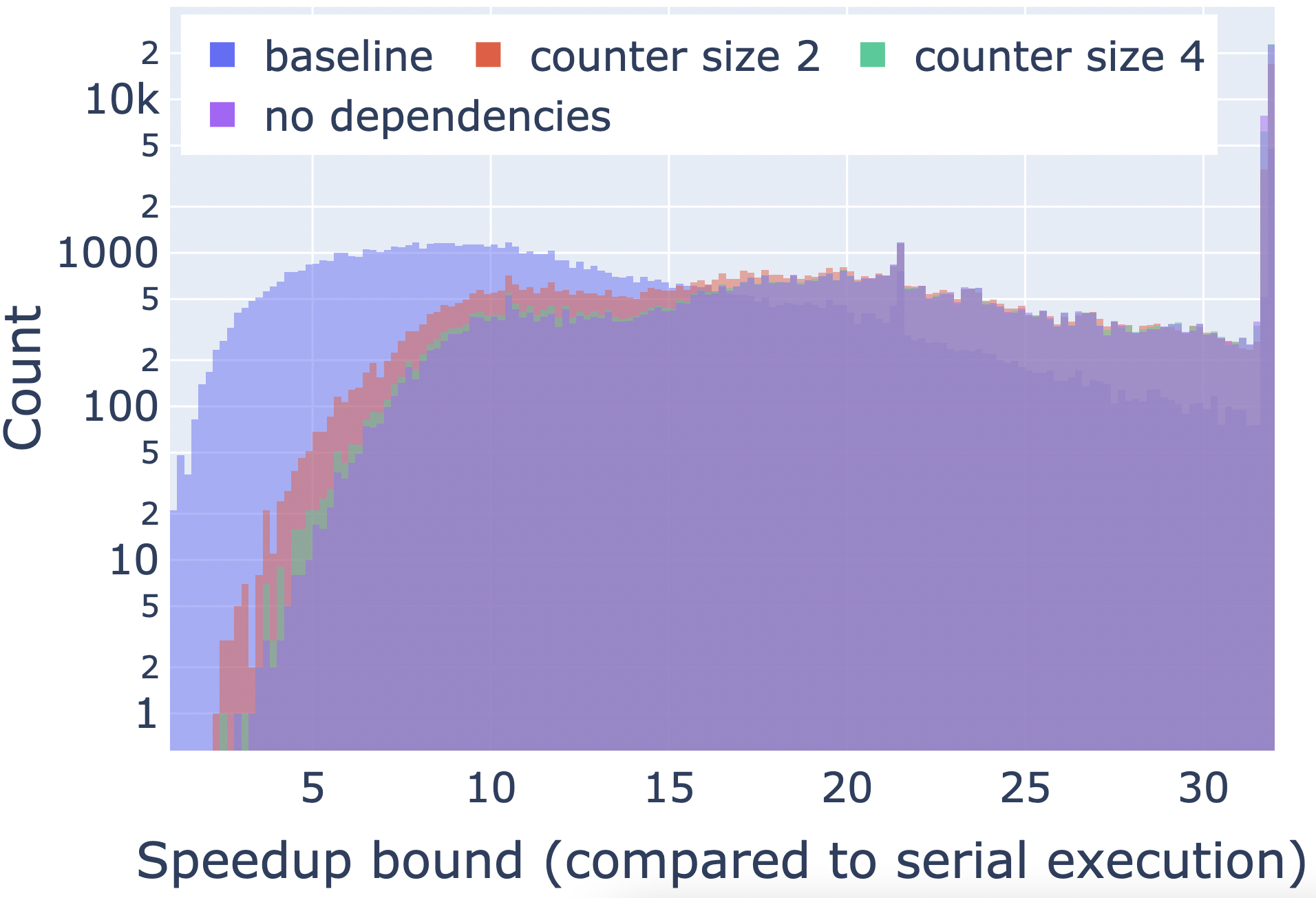}
    \caption{Distribution of speedup bounds (10-block batches)}
    \label{fig:partitioned_hist}
\end{figure}

\begin{figure}
    \centering
    \includegraphics[width=0.3\textwidth]{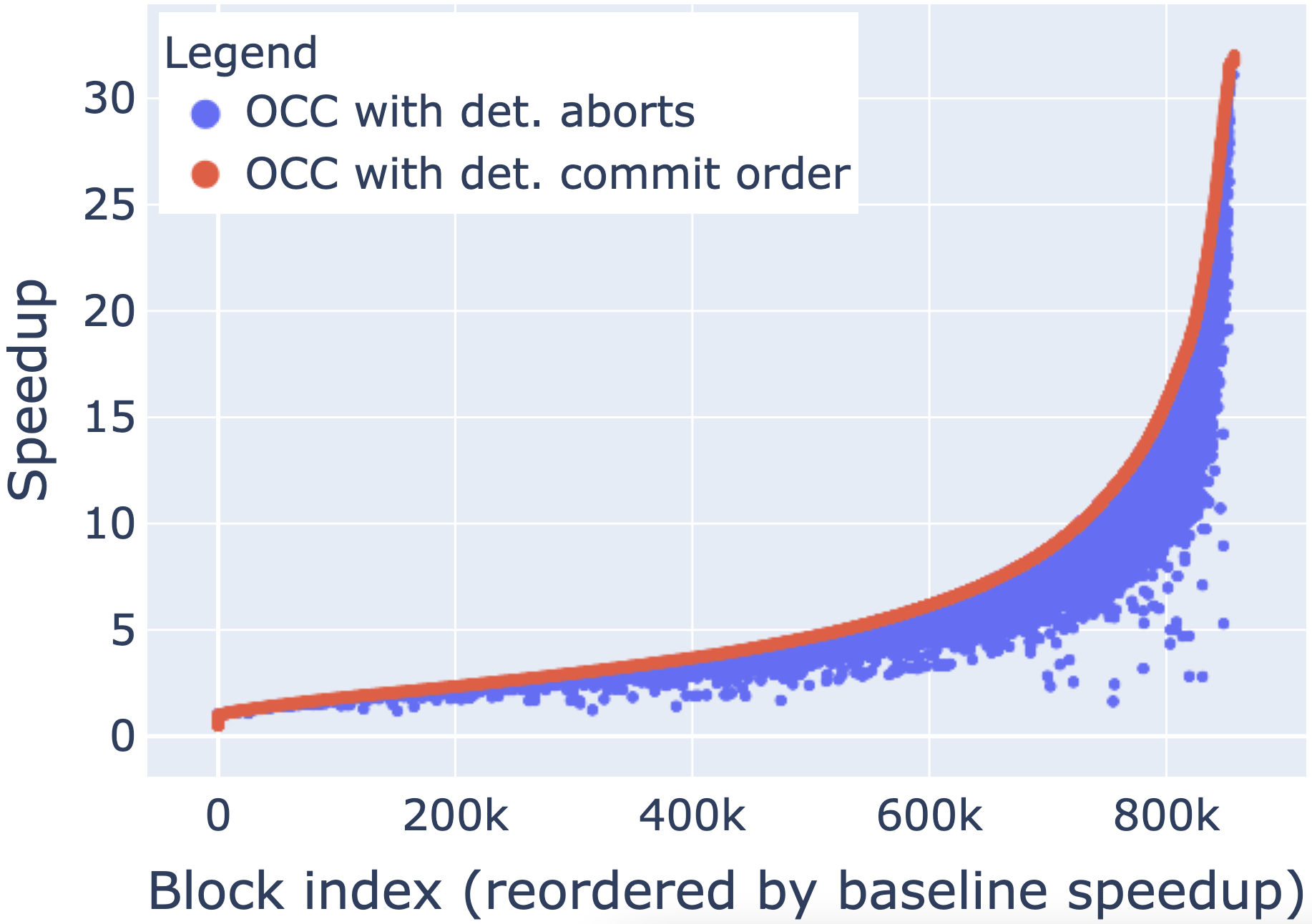}
    \caption{OCC-DA performance impact}
    \label{fig:deter_deg}
\end{figure}

As for OCC-DA, over single blocks with 32 threads, the baseline OCC scheduler has a 3.287x overall speedup (min: 0.52x, max: 32x, avg: 5.89x), while the deterministic scheduler results in 3.275x overall speedup (min: 0.52x, max: 32x, avg: 5.84x). We observed similar results over batches of 30 blocks.

\paragraph{Discussion.} These results show that the parallelism inherent in the dataset (9.25x) is much lower than what the transactions would allow for (23.63x). This is due to the fact that transactions depending on each other need to be scheduled serially (or get aborted). By eliminating some dependencies using techniques like partitioned counters, we can approach this limit, achieving up to 17.96x speedup with just a counter size of 2. Figure \ref{fig:partitioned_hist} shows overlayed histograms for the distribution of speedup bounds for each block-batch. From this figure, we can clearly see how partitioned counters let us converge to the optimum, in terms of the parallel speedup achievable.

Figure \ref{fig:deter_deg} shows the performance degradation caused by OCC-DA (blue), compared to OCC with det. commit order (red) on single blocks. For this figure, blocks were ordered by their baseline speedup (red). We can see that extending the scheduler with deterministic aborts did cause performance degradation, however, the speedups generally still do not diverge much from the baseline, except for a few outliers. In fact, in this dataset, 92.47\% of the blocks produced exactly the same result using the two schedulers, while only 0.22\% resulted in 80\% of the baseline speedup or lower.

\paragraph{Implications.} These results suggest that partitioned counters can have a significant impact on the highest parallel speedup that we can achieve. Even with just a counter of length 2, when applied to all conflicts, the parallel speedup bound doubled, approaching the optimum. Raising the counter length, we keep approaching the optimum. Based on these results, we believe that the techniques proposed in this paper, when applied to some contracts responsible for some major bottlenecks, can significantly increase the parallel speedup that any real-world parallel scheduler can achieve.

The results about OCC-DA suggest that raising the level of determinism only has a minor performance impact, decreasing the overall speedup from 3.287x to 3.275x. As shown in Figure  \ref{fig:deter_deg}, while there are occasional outliers with significant performance degradation under this scheduling mechanism, they are rare. While it is possible that a more performant scheduler, and a workload with more parallelism, will result in a larger discrepancy between these two numbers, based on these initial evaluations, our expectation is that OCC-DA is suitable for implementation in real-world blockchain protocols.
\section{Threats to Validity} \label{threats_to_validity}

The most significant threat to the validity of our study is that transaction and contract interaction patterns have changed since the observed period in 2018 and so our conclusions do not hold for more recent periods. We believe that this is unlikely. The issues pointed out in this paper have not been addressed, and so there has been neither awareness nor incentive to avoid these conflict-inducing patterns. If anything, the problem has likely become more severe, with several new hotspot contracts emerging, many of which have obvious storage conflicts. An example for this is Uniswap, as pointed out in Section \ref{sec:generalizability}. Saraph et al.~\cite{saraph2019empirical} also observes that the parallelizability of blocks seems to decrease over time.

There is a chance that the gas cost of transactions does not accurately capture their running time, which would reduce the accuracy of our evaluations. Given that the most time-consuming operations (namely, \texttt{SLOAD} and \texttt{SSTORE}) have very high gas costs, large deviation seems unlikely.

In this study, we only considered storage conflicts. Other conflict types include conflicts on an account's balance and nonce, and conflicts on contract creation/destruction. Balance conflicts can be handled using partitioned counters. Nonce conflicts require adjusting the nonce management mechanism. As contract existence conflicts are rare, they are unlikely to have distorted our results.
\section{Relevance and Future Work} \label{discussion}

\paragraph{Scope.} Our work focuses on Ethereum but our findings and solutions are applicable to other chains as well. This claim is supported by two trends. First, many significant blockchains adopt Ethereum's execution logic (Ethereum Classic, BSC), while others are in the process of adding an EVM compatibility layer (Near, Solana). Second, different chains share common use cases (DeFi swap platforms, NFT marketplaces), and popular Ethereum applications are often forked and redeployed on other chains (Uniswap and PancakeSwap). This results in similar transaction workloads on these chains. This similarity is supported by previous research as well \cite{reijsbergen2020exploiting}.

\paragraph{Feasibility.} Implementing parallel execution of blockchain transactions requires further research and engineering effort. We need further research on parallel execution incentives, and the EVM needs to be extended with a parallel scheduler. OCC-DA provides a foundation for these by offering a parallel scheduling framework suitable for distributed consensus protocols. We also need techniques for increasing the parallelizability of the transaction workload. The proposed bottleneck-elimination techniques address this requirement. Of these techniques, only \emph{"Commutative EVM Instructions"} requires a protocol upgrade. Finally, further work is required to design an optimized deterministic parallel scheduler, building on existing techniques from traditional database systems.
\section{Related Work} \label{related_work}

Parallel execution of blockchain transactions has been the focus of considerable research attention in recent years.
Perhaps the first such work is by Sergey et al.~\cite{sergey2017concurrent}, in which the authors propose to treat smart contracts as concurrent objects to prevent common bugs. In 2019, Saraph et al.~\cite{saraph2019empirical} published an exploratory work to estimate the potential benefit of executing Ethereum transactions in parallel by simulating a 2-phase parallel-then-serial optimistic scheduler. They observe a 2-fold speedup for the period in 2018 using 64 threads, and identify \textit{CryptoKitties} as a hotspot contract. They briefly remark on incentives and commutative operations. Reijsbergen et al.~\cite{reijsbergen2020exploiting} evaluate the potential speedup on seven public blockchains using dependency graphs, working on the granularity of contracts instead of storage entries. They report that up to 6x speedup is achievable using 8 or more cores, and observe that larger blocks are easier to parallelize. Our empirical study is inspired by these two works, and we reinforce or expand on some of their conclusions. However, these works do not analyze conflicts in-depth and so they fail to explain the poor parallel speedups they predict. Their models also do not fulfill the determinism requirements that would make them practical in public blockchains.

Numerous previous works have proposed to use various concurrency control techniques to parallelize blockchain transactions.
In the approach proposed by Anjana et al.~\cite{anjana2019efficient}, miners use optimistic STM to execute transactions and produce a dependency graph that validators can use to re-execute transactions.
Zhang et al.~\cite{zhang2018enabling}, instead of using a dependency graph, propose to include each transaction's write set in the block, and let validators use these to detect conflicts.
Pang et al.~\cite{pang2019concurrency} also consider the granularity of the additional information included in the block.
Dickerson et al.~\cite{dickerson2019adding} propose to use abstract locks to detect conflicts during speculative parallel execution. Dozier et al.~\cite{doziercorrectness}, on the other hand, use a Pessimistic Concurrency Control technique by locking the accounts accessed during transaction execution.
Finally, Bartoletti et al.~\cite{bartoletti2020true} offer a formal model of concurrent blockchain transactions.
Most of the proposed techniques are protocol-breaking, in the sense that they modify the block structure and the execution semantics, while our approach remains compatible with serial implementations. These works show modest speedup on parallel miners but they do not address the root cause of the speedup limit.
An overview of this area can be found in the survey piece by Kemmoe et al.~\cite{kemmoe2020recent}.

Optimistic concurrency control~\cite{kung1981optimistic} has been widely used in databases and wide-area distributed systems.
Deterministic OCC was pioneered by Abadi et al.
In Calvin~\cite{thomson2012calvin}, they use a deterministic locking protocol to let nodes arrive on a consistent transaction order, eliminating the need for distributed commit protocols.
Their approach is further outlined in several other papers~\cite{thomson2010case, thomson2011building, abadi2018overview}.
Our discussion of the determinism of blockchain transaction execution was inspired by these works.
In addition to using a predefined serialization order, we introduced an even higher level of determinism, where the effects of transactions that are normally not observable are also deterministic, and can be used for incentive assignment.

\vspace{-3mm}

\section{Conclusion} \label{conclusion}

With the evolution of consensus protocol technology in public blockchain, the execution efficiency is becoming the new bottleneck of the entire system, which drives the need of parallelizing the transaction execution.
This work observes that the application inherent conflicts are the fundamental obstacle to achieving ideal speedup in existing parallelization techniques.
To address this issue, the proposed solution introduces the convenient improvement on the smart contract programming paradigm with consideration of the support of incentives, therefore opens the possibility of maximizing the parallelism of transaction execution in public blockchains.

\bibliographystyle{ACM-Reference-Format}
\bibliography{bibliography}

\end{document}